\documentclass{aa}
\usepackage{graphicx}
\begin{document}
 
\title{A statistical method to determine open cluster metallicities}
 
\author{H.~P{\"o}hnl and E.~Paunzen}


\institute{Institut f{\"u}r Astronomie der Universit{\"a}t Wien, T{\"u}rkenschanzstr. 17,
A-1180 Wien, Austria\\
\email{Ernst.Paunzen@univie.ac.at}}

\date{}

\abstract
{The study of open cluster metallicities helps to understand the local
stellar formation and evolution throughout the Milky Way. 
Its metallicity gradient is an important tracer for the Galactic
formation in a global sense. Because open clusters can be treated in
a statistical way, the error of the cluster mean is minimized.}
{Our final goal is a semi-automatic statistical robust method 
to estimate the metallicity of a statistically significant number of 
open clusters based on Johnson $BV$ data of their members,
an algorithm that can easily be extended to other photometric systems
for a systematic investigation.}
{This method incorporates evolutionary grids
for different metallicities and a calibration of the 
effective temperature and luminosity. With cluster
parameters (age, reddening and distance) it is possible to estimate
the metallicity from a statistical point of view. The iterative
process includes an intrinsic consistency check of the starting input 
parameters and allows us to modify them.
We extensively tested the method with published data for the Hyades.}
{We selected sixteen open clusters within 1000\,pc around the Sun
with available and reliable Johnson $BV$ measurements. In
addition, Berkeley 29, with a distance of about 15\,kpc was chosen.
For several targets we are able to compare our result with published ones
which yielded a very good coincidence (including Berkeley 29).}
{A new method for the statistical determination of open 
cluster metallicities is presented and tested. It is quite
robust against 
errors in effective temperature and luminosity calibration of the individual stars.}
\maketitle
 
\section{Introduction}

Open clusters comprise of a local stellar population of with the
same age and metallicity at a distinct distance from the Sun. Their
study provides valuable information on the local as well as the global
environment. In general, determining the age, distance and reddening for
an open cluster by fitting isochrones is rather straightforward. However,
the fourth free parameter, the metallicity, is either set to the solar
value or is simply neglected.

\begin{table*}[ht]
\begin{center}
\caption[]{Distance, reddening and age starting values for the seventeen open clusters
used in our analysis are presented in the left panel. In the right panel
are the results obtained with our analysis.}
\begin{tabular}{llrcc|rccccl}
\hline
\hline
Cluster & & \multicolumn{1}{c}{$d$} & $E(B-V)$ & log\,$t$ & \multicolumn{1}{c}{$d$} & $E(B-V)$ & log\,$t$ & [Z] & [Fe/H] & \multicolumn{1}{c}{[Fe/H]$_\mathrm{L}$} \\
& & [pc] & [mag] & [dex] & [pc] & [mag] & [dex] & [dex] & [dex] & \multicolumn{1}{c}{[dex]} \\
\hline
Alessi 13 & & 110 & 0.040 & 8.720 & 110 & 0.040 & 8.720 & 0.027(7) & +0.17 & \\
Berkeley 29 & C0650+169 & 14870 & 0.157 & 9.025 & 16440 & 0.200 & 9.380 & 0.007(2) & $-$0.48 & $-$0.31(3)$^2$ \\
IC 4665 & C1743+057 & 352 & 0.174 & 7.634 & 352 & 0.174 & 7.634 & 0.022(9) & +0.08 & $-$0.03(4)$^2$ \\
Melotte 20 & Alpha Per & 185 & 0.090 & 7.854 & 185 & 0.090 & 8.050 & 0.028(7) & +0.18 & $-$0.05$^1$ \\
Melotte 25 & Hyades & 45 & 0.010 & 8.896 & 46 & 0.010 & 8.900 & 0.028(7) & +0.18 & +0.13(5)$^2$ \\
Melotte 111 & Coma Ber & 96 & 0.013 & 8.652 & 81 & 0.013 & 8.800 & 0.018(8) & $-$0.03 & $-$0.05$^1$ \\
NGC 752 & C0154+374 & 457 & 0.034 & 9.050 & 429 & 0.044 & 9.100 & 0.021(5) & +0.05 & $-$0.08$^1$ \\
NGC 1039 & C0238+425 & 499 & 0.070 & 8.249 & 499 & 0.070 & 8.250 & 0.023(7) & +0.10 & $-$0.30$^1$ \\
NGC 2168 & C0605+243 & 816 & 0.262 & 7.979 & 816 & 0.262 & 8.200 & 0.014(3) & $-$0.15 & $-$0.16$^1$ \\
NGC 2451\,A & C0743$-$378 & 189 & 0.010 & 7.780 & 189 & 0.010 & 7.780 & 0.020(5) & +0.02 & \\
NGC 2451\,B & C0743$-$378 & 302 & 0.055 & 7.648 & 302 & 0.055 & 7.570 & 0.019(4) & $-$0.01 & \\
NGC 2516 & C0757$-$607 & 409 & 0.101 & 8.052 & 360 & 0.112 & 8.150 & 0.015(5) & $-$0.12 & +0.06$^1$ \\
NGC 2547 & C0809$-$491 & 455 & 0.041 & 7.557 & 382 & 0.060 & 7.500 & 0.017(5) & $-$0.06 & $-$0.16$^1$ \\
NGC 2632 & Praesepe & 187 & 0.009 & 8.863 & 171 & 0.027 & 8.720 & 0.028(6) & +0.18 & +0.14$^1$ \\
NGC 6475 & C1750$-$348 & 301 & 0.103 & 8.475 & 301 & 0.103 & 8.300 & 0.028(7) & +0.18 & +0.14(6)$^2$ \\
NGC 7092 & C2130+482 & 326 & 0.013 & 8.445 & 326 & 0.013 & 8.445 & 0.026(6) & +0.15 & \\
Stock 2 & C0211+590 & 303 & 0.380 & 8.230 & 380 & 0.340 & 8.230 & 0.032(10) & +0.25 & \\
\hline
\multicolumn{11}{l}{1: Chen et al (2003), 2: Magrini et al. (2009)}
\end{tabular}
\label{targets_webda}
\end{center}
\end{table*}

But the intrinsic metallicity of (proto-)stars is an important 
key parameter for our understanding of stellar formation and evolution. 
Metallicity already severely influences the cooling and collapse of ionized gas 
during the first stages of star formation (Jappsen et al. 2007). Machida (2008)
showed that clouds with lower metallicity have a higher probability of 
fragmentation, indicating that the binary frequency is a decreasing 
function of cloud metallicity. On larger scales, clusters form and evolve
significantly different in different environments, for example in the Milky
Way and the Magellanic Clouds (Johnson et al. 1999). 

Investigating the global properties of our Milky Way, a metallicity gradient 
throughout the Galactic disk was discovered several decades ago. 
The study of this gradient provides strong
constraints on the mechanism of galaxy formation. Models show that star
formation and the accretion history as functions of galactocentric
distance strongly influence the appearance and
the development of the abundance gradients (Chiappini et al. 2001).
Relevant studies either use stellar data, for example those of Cepheids 
(Andrievsky et al. 2004, Cescutti et al. 2007), or open clusters 
(Chen et al. 2003, Magrini et al. 2009) as well as globular clusters 
(Yong et al. 2008).
The first approach is limited to the accurate distance estimation of field
stars and the uncertainties of spectroscopic abundance analysis for very
distant and therefore faint objects,
whereas 
metallicity compilations of open clusters (for example Chen et al. 2003) are 
normally based on rather inhomogeneous data sets. 

We present a statistical method to estimate the metallicity of an open 
cluster based on photometric data sets; mostly on Johnson $BV$, for
brighter stars also Geneva and Str{\"o}mgren; and isochrone fitting to its
members with an approximate knowledge about the clusters' 
age, reddening and distance. The latter three parameters can be derived 
independently from different photometric data sets (Mermilliod \& Paunzen 2003) 
or can be taken from catalogues (Paunzen \& Netopil 2006).

The presented method can be easily extended to any other photometric
system for which isochrones are available. This is especially interesting for
all-sky-surveys like 2MASS (Cutrie et al. 2003) including homogeneous 
data of open clusters. Furthermore, it includes an intrinsic 
consistency check of the starting input parameters, hence it is superior to a standard
isochrone fitting technique.

In Sect. \ref{amts} we present the basic method, its application to the
Hyades and the target selection.  
In Sect. \ref{analysis} we apply our method to sixteen additional open clusters. 
Several of them are compared to previous, independent investigations.

\section{Aims, methods and target selection} \label{amts}

Recently, Magrini et al. (2009) published metallicities for 45 open clusters on the 
basis of high resolution spectroscopy from the literature (mainly from red giants). 
However, it is still unclear whether the elemental abundances
derived from highly evolved members of open clusters really represent those
of stars still on the Main Sequence (MS hereafter). Another project in this respect is
the Bologna Open Cluster Chemical Evolution
project (Bragaglia 2008), dedicated primarily to old open clusters.
Such surveys for open clusters are rather time consuming. 

Our final goal is a semi-automatic statistically robust method 
to estimate the metallicity of a statistically significant sample
of open clusters based on 
Johnson $BV$ data of its members. 
The method is designed in a way that it can easily be extended
to other photometric systems like
$VI_\mathrm{c}$. The influence of calibration errors of the
basic astrophysical parameters for the individual stars should 
be minimized.

The procedure incorporates published evolutionary grids
for different metallicities (Sect. \ref{grids}) and a
calibration for the effective temperature and luminosity
(Sect. \ref{eff_lum}).

Targets were selected on the basis of well-known cluster
parameters (distance, reddening and age) within 1000\,pc. 
Furthermore, available Johnson $BV$ photometry 
to define a reliable MS for each cluster 
was another criterion. In order to test our method also
at larger galactic distances, we additionally chose 
Berkeley~29 (about 15\,kpc away from the Sun). For this old cluster,
widely different cluster parameters are published. 
In total, we investigated seventeen open clusters 
in WEBDA\footnote{http://www.univie.ac.at/webda} that fulfil
our requirements (Table \ref{targets_webda}). They are evenly
distributed according to the reddening and age values.

The consecutive steps of our method, incorporating 
a test using the data of the Hyades are
described in Sect. \ref{test_hyades}.

In the literature, metallicities are
listed either as [Fe/H] or [Z] values. If not indicated
otherwise, these parameters can be transformed using 
[Y]\,=\,0.23\,+\,2.25[Z] derived by Girardi et al. (2000).
For clarity, we list the corresponding values used for this
paper in Table \ref{fe_z}.

For the remainder of the paper the errors in the final digits 
of the corresponding quantity are given in parenthesis.

\begin{table}
\begin{center}
\caption[]{Transformation of [Fe/H] to [Z].}
\begin{tabular}{cccccc}
\hline
\hline
[Fe/H] & [Z] & [Fe/H] & [Z] & [Fe/H] & [Z] \\
\hline
$-$0.729 & 0.004 & $-$0.030 & 0.018 & +0.253 & 0.032 \\ 
$-$0.525 & 0.006 & +0.019 & 0.020 & +0.288 & 0.034 \\
$-$0.387 & 0.008 & +0.077 & 0.022 & +0.312 & 0.036 \\
$-$0.282 & 0.010 & +0.116 & 0.024 & +0.343 & 0.038 \\
$-$0.224 & 0.012 & +0.152 & 0.026 & +0.371 & 0.040 \\
$-$0.149 & 0.014 & +0.185 & 0.028 \\
$-$0.086 & 0.016 & +0.225 & 0.030 \\
\hline
\end{tabular}
\label{fe_z}
\end{center}
\end{table}

\subsection{The grid of evolutionary models} \label{grids}

Our intention is to determine the metallicity on the basis
of a theoretical Hertzsprung-Russell diagram, 
that means basically via
$\log L/L_{\sun}$ versus $\log T_\mathrm{eff}$. 
Below we use the classical notation
[H,He,Others] for [X,Y,Z] and/or [X:Y:Z]. The starting point
is the evolutionary grid published by Claret \& Gimenez (1998) 
available for [X:Z]\,=\,[0.63,0.73,0.80:0.01],
[0.60,0.70,0.80:0.02] and [0.55,0.65,0.75:0.03].
The models allow us to take into account a known Helium abundance for
open clusters (e.g. Hyades). Furthermore, they are very densely tabulated, 
justifying a linear interpolation. For 
[X:Y:Z]\,=\,[0.744:0.252:0.004] as well as [0.62:0.34:0.04], 
which are not available by Claret \& Gimenez (1998), we used the models 
published by Schaller et al. (1992), Charbonnel et al. (1993) 
and Schaerer et al. (1993). The mentioned three grids 
are consistent within each other.

As a first step, the compatibility of the above mentioned grids 
was tested. For this purpose, the Zero-Age-Main-Sequence (ZAMS hereafter)
of the model [X:Y:Z]\,=\,[0.68:0.30:0.02] was calculated and compared to
the others.
For the range 0\,$\le$\,$\log L/L_{\sun}$\,$\le$\,3, with differences for 
$\log T_\mathrm{eff}$ between 0.006 and 0.010\,dex which corresponds to $\le$\,2\% 
in $T_\mathrm{eff}$ only.

Table \ref{logteff_xc} shows the dependence of $\log T_\mathrm{eff}$ to
[X:Y] for a constant [Z] value. In general, a mean [Y] value is assumed,
which is related to the solar one. We were able to study published [Y]
abundances for the case of the Hyades as described in 
Sect. \ref{test_hyades}. Perryman at al. (1998) derived
[Y]\,=\,0.26(2), whereas de Bruijne et al. (2001) list 
[Y:Z]\,=\,[0.285:0.024], taking into account new results for the Sun. 
The latter values agree with the Y-Z-relation published by 
Girardi et al. (2000). An uncertainty of a few percent for [Y] corresponds
to only a small alteration of $\log T_\mathrm{eff}$ (Table \ref{logteff_xc}), 
which is well in the range of the ``grid differences''. 
An absolute error of 3\% in the temperature determination ($\approx$\,200\,K) 
is equal to $\Delta$[Y]\,=\,0.1, which is well beyond the observed abundance
uncertainties.

With [Y]\,=\,0.23\,+\,2.25[Z] as published by Girardi et al. (2000), we chose
the following models for our calibrations: [0.744:0.252:0.004], [0.73:0.26:0.01], 
[0.70:0,28:0.02], [0.65:0.32:0.03], and [0.62:0.34:0.04]. 

The final grid used for our analysis is listed in Table \ref{logteff_grids1} in
time steps of $\Delta$log\,$t$\,=\,0.2 for the complete metallicity and age
range from 7.2\,$\le$\,$\log t$\,$\le$9.6.

\begin{figure}
\begin{center}
\includegraphics[width=80mm]{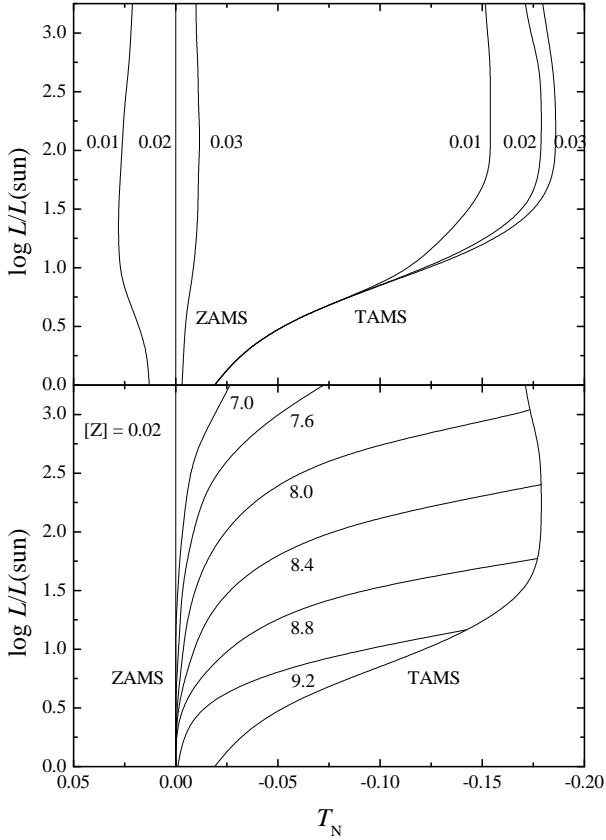}
\caption{The evolutionary grid for different metallicities and ages versus the normalized
effective temperature ($T_\mathrm{N}$).}
\label{plot_grids}
\end{center}
\end{figure}
\begin{table}
\begin{center}
\caption[]{The ZAMS (left panel) and TAMS (right panel)
for [Z]\,=\,0.02 and different core Hydrogen abundances ($X_\mathrm{c}$).}
\begin{tabular}{cccc|ccc}
\hline
\hline
$X_\mathrm{c}$ & 0.6 & 0.7 & 0.8 & 0.6 & 0.7 & 0.8 \\
\hline
$\log L/L_{\sun}$ & \multicolumn{3}{c|}{$\log T_\mathrm{eff}$} 
& \multicolumn{3}{c}{$\log T_\mathrm{eff}$} \\
0.3 & 3.815 & 3.798 & 3.785 & 3.798 & 3.780 & 3.764 \\
1.2 & 3.967 & 3.953 & 3.938 & 3.829 & 3.803 & 3.785 \\
2.1 & 4.131 & 4.118 & 4.107 & 3.975 & 3.939 & 3.902 \\
3.0 & 4.289 & 4.277 & 4.266 & 4.142 & 4.103 & 4.065 \\
\hline
\end{tabular}
\label{logteff_xc}
\end{center}
\end{table}

\subsection{Calibration of the effective temperature and luminosity} \label{eff_lum}

For the calibration of the luminosity, an estimation for the bolometric correction
$BC$ is vital. We took the results of Flower (1996), who lists $BC$ versus $\log T_\mathrm{eff}$
and ($B-V$) for MS stars and Giants. His Table 3 includes values for
$-$0.35\,$<$\,($B-V$)\,$<$\,+1.8, which can be immediately compared to the calibration
by Alonso et al. (1996) for [Fe/H]\,=\,0. The differences $\Delta\log T_\mathrm{eff}$ between these
two lists for +0.04\,$<$\,($B-V$)\,$<$\,+0.64 and the complete range of metallicities
is less than 1\%.

A polynomial interpolation of the data of Flower (1996) was performed, resulting in the following
equations
\begin{eqnarray}
BC &=& -0.238 - 5.561(\log T_\mathrm{eff} - 4) \label{bc1} \\ 
&& \qquad \qquad \qquad \qquad \mbox{for}\,\,4.00 < \log T_\mathrm{eff} < 4.25 \nonumber \\
BC &=& -0.243 - 5(\log T_\mathrm{eff} - 4) - 26(\log T_\mathrm{eff} - 4)^2 - \nonumber \\
&& -33.33(\log T_\mathrm{eff} - 4)^3 \\
&& \qquad \qquad \qquad \qquad \mbox{for}\,\,3.76 < \log T_\mathrm{eff} < 4.00 \nonumber \\
BC &=& -0.155 + 2.84(B-V) - 15.13(B-V)^2 + \nonumber \\
&& +25.57(B-V)^3 \\
&& \qquad \qquad \qquad \qquad \mbox{for } -0.2 < (B-V) < +0.1 \nonumber \\
BC &=& -0.030 + 0.491(B-V) - 0.937(B-V)^2 + \nonumber \\
&& +0.055(B-V)^3 \label{bc4} \\
&& \qquad \qquad \qquad \qquad \mbox{for } +0.1 < (B-V) < +0.8, \nonumber \\
\end{eqnarray}
which are used to determine the $BC$. Finally the luminosity can be derived, taking 
$M_\mathrm{\sun,Bol}$\,=\,4.74\,mag from Bessell et al. (1998) into account as
\begin{equation}
\log L/L_{\sun} = 1.896 - 0.4(M_\mathrm{V} + BC) \label{bc_flower}
\end{equation}
within the listed range of parameters.

For establishing a metallicity calibration of an open
cluster without narrow band and/or reddening independent photometry,
we needed a well established and tested effective temperature calibration for
($B-V$). Alonso et al. (1996) published a $\log T_\mathrm{eff}$--[Fe/H]--($B-V$)
calibration valid for +0.2\,$<$\,($B-V$)\,$<$\,+1.5 (F0 to K5) and
$-$0.5\,$<$\,[Fe/H]\,$<$\,+0.5 (suitable for open clusters) as follows
\begin{eqnarray}
\Theta_\mathrm{eff} = &+&0.541 + 0.533(B-V) + 0.007(B-V)^2 - \nonumber \\
&-&0.019(B-V)[\mathrm{Fe/H}] - 0.047[\mathrm{Fe/H}] - \nonumber \\
&-&0.011[\mathrm{Fe/H}]^2, \label{A96_1}
\end{eqnarray}
where $\Theta_\mathrm{eff}$\,=\,5040/$T_\mathrm{eff}$.
We checked the results of the calibration with the values for Hyades members
listed in Perryman at al. (1998) and found the following statistically significant
correction term (see Sect. \ref{test_hyades})
\begin{equation}
\log T_\mathrm{eff}^\mathrm{corr} - 
\log T_\mathrm{eff} = -0.009 + 0.012(B-V), \label{A96_2}
\end{equation}
In the paper by Alonso et al. (1996), several comparisons with other calibrations
can also be found.

For further interpolation we used normalized logarithmic effective temperature $T_\mathrm{N}$ values
defined as
\begin{equation}
T_\mathrm{N} = \log T_\mathrm{eff}(\log L/L_{\sun}) -
\log T_\mathrm{eff}(\log L/L_{\sun})_\mathrm{ZAMS(Z=0.02)}, \label{tn}
\end{equation}
which simplified the calculations significantly. 
The basis of the normalization were the values of the ZAMS model for [0.70:0.28:0.02]
as listed in Table \ref{list_norm}. In other words, the evolution from the ZAMS for stars results
always in a lower $T_\mathrm{N}$. If one intends to use other isochrone models, only the normalization
ZAMS has to be altered.
Figure \ref{plot_grids} shows the grid for solar metallicity and 
different ages (lower panel). The upper panel shows the dependence of the ZAMS and Terminal-Age-Main-Sequence
(TAMS hereafter)  on the
metallicity. The standard lines of $T_\mathrm{N}$ are much better separated for the metal poor 
than for the metal rich models. The differences for the region close to the TAMS are less pronounced.
It is therefore essential to provide a profound knowledge about the cluster MS.

\begin{table}
\begin{center}
\caption[]{The ZAMS of the evolutionary model used for the normalization.}
\begin{tabular}{cc|cc}
\hline
\hline
$\log L/L_{\sun}$ & $\log T_\mathrm{eff}$ & $\log L/L_{\sun}$ & $\log T_\mathrm{eff}$ \\
\hline
$-$0.3 & 3.725 & +1.8 & 4.064 \\
+0.0 & 3.766 & +2.1 & 4.118 \\
+0.3 & 3.798 & +2.4 & 4.171 \\
+0.6 & 3.835 & +2.7 & 4.225 \\
+0.9 & 3.893 & +3.0 & 4.277 \\
+1.2 & 3.953 & +3.3 & 4.327 \\
+1.5 & 4.009 & & \\
\hline
\end{tabular}
\label{list_norm}
\end{center}
\end{table}

\subsection{Test case: the Hyades} \label{test_hyades}

The Hyades were a natural choice for testing our method because they
are well investigated and their metallicity is far from being settled.
If our method is not robust at all, the result should be far off from
the published values because the astrophysical parameters of the members 
are well known. In addition, we would like to point out that Melotte
20, Melotte 111, NGC 2632, and NGC 6475 are also included in this study. 
These open clusters are, in general, taken as ``standards'' in various
papers.
The main source of our investigation is the paper by 
Perryman at al. (1998), who extensively discussed the membership
probabilities, distance, age and metallicity. They found a best 
isochrone fit with the model [0.716:0.26:0.024]. 
Here is an incomplete review of published [Fe/H] values
\begin{itemize}
\item Cameron (1985b): +0.08, photometry
\item Berthet (1990): +0.081, photometry
\item Boesgaard \& Friel (1990): +0.127(22), high resolution spectroscopy 
\item Varenne \& Monier (1999): $-$0.05(15), high resolution spectroscopy
\item Dias et al. (2002): +0.17, mean value from publications
\item Paulson et al. (2003): +0.13(1), high resolution spectroscopy 
\item Taylor \& Joner (2005): +0.103(8), photometry and spectroscopy
\end{itemize}
showing the overall range of the estimates.

In Table 8 by Perryman at al. (1998), 40 Hyades members are listed, for which
high resolution spectroscopy, photometry, parallaxes, effective temperatures,
bolometric magnitudes and abundances are available. From this list, we excluded 
known spectroscopic binaries, chemically peculiar stars and giants which are 
too far from the ZAMS, 
which left us with 33 objects. These give a distance modulus of 3.33(1)
or 46\,pc for the cluster centre. The members are within
10\,pc around the centre and necessitate the use of the individual distances for deriving 
the luminosities.

The reddening was set to 0.01\,mag, which is an upper limit for the Hyades 
(de Bruijne et al. 2001). Any smaller value has no significant influence
on the effective temperature calibration. The age of the Hyades is estimated as
625(50)\,Myr by Perryman at al. (1998), whereas An et al. (2007) give
550\,Myr. In Dias et al. (2002) we find a significant higher value of 787\,Myr.

\begin{figure}
\begin{center}
\includegraphics[width=80mm]{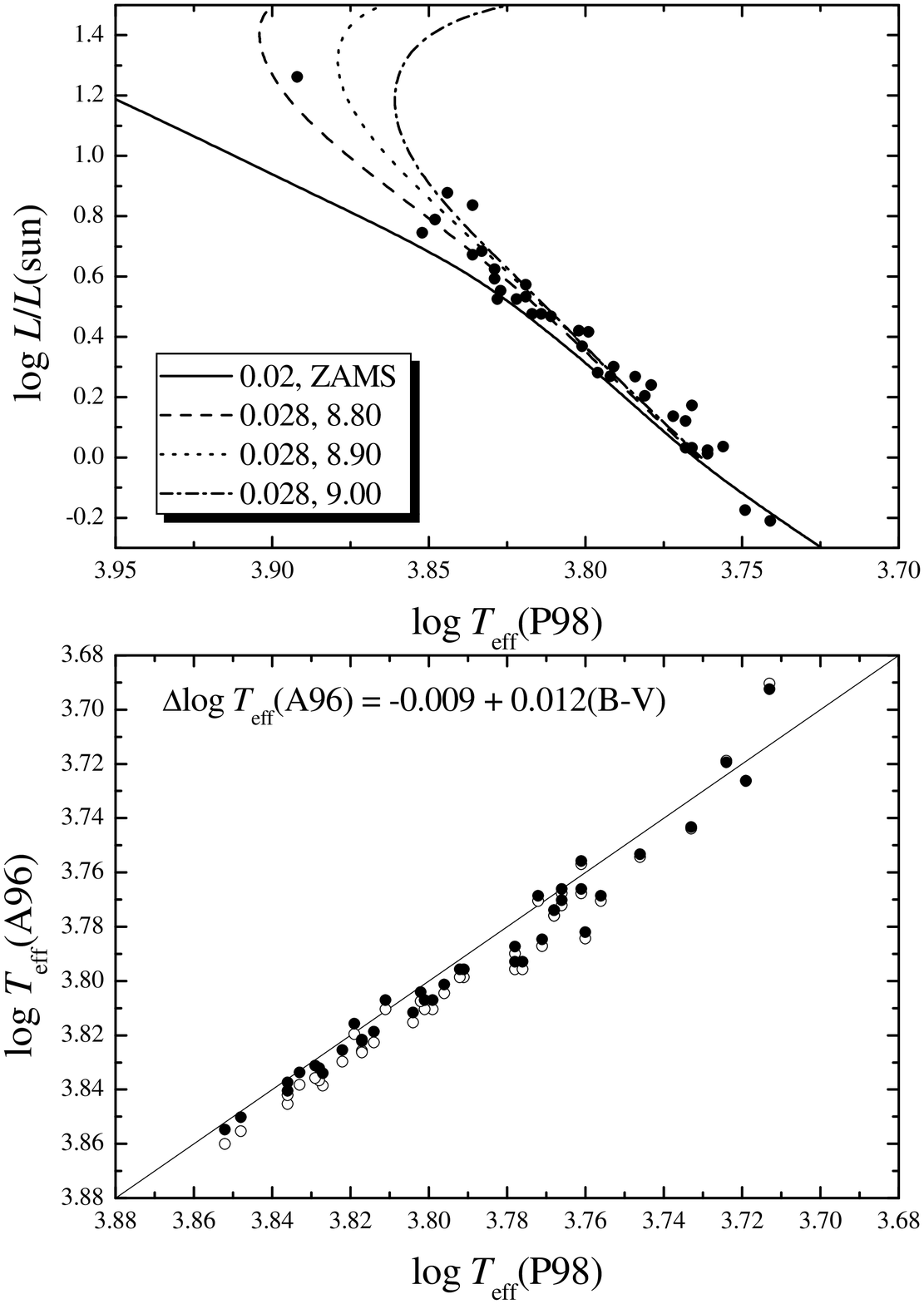}
\caption{The isochrones for the Hyades with the members taken from Perryman at al. (1998, P98). 
The lower panel shows the offset to the calibration of Alonso et al. (1996, A96).}
\label{plot_hyades}
\end{center}
\end{figure}

The published metallicity can be estimated by the weights [0.60,0.36,0.04]
of the following isochrones [0.70:0.28:0.02], [0.75:0.22:0.03] and [0.65:0.32:0.03],
respectively. Figure \ref{plot_hyades} shows the isochrones for the Hyades
with the apparent members. The lower panel shows the
offset to the calibration of Alonso et al. (1996) according to Sect. \ref{eff_lum} with
(filled circles) and without (open circles) correction. According to this figure,
we deduce log\,$t$\,=\,8.90 for our further analysis. The mean difference 
$\Delta\log T_\mathrm{eff}$ of the calibrated to the isochrone value 
for stars with $\log L/L_{\sun}$\,$>$\,0 is $-$0.002(7).

Below we list the consecutive steps of the iterative process 
of our method, the input parameters are $V$, $(B-V)$ and the grid 
$\log L/L_{\sun}$-$T_\mathrm{N}$-[Fe/H] (Table \ref{logteff_grids1})
\begin{enumerate}
\item $\log T_\mathrm{eff}$ = f($(B-V)_\mathrm{0}$,[Fe/H]) [Eq. \ref{A96_1}]
\item BC = f($(B-V)_\mathrm{0}$,$\log T_\mathrm{eff}$) [Eq. \ref{bc1} - \ref{bc4}] 
\item $\log L/L_{\sun}$ = f($M_\mathrm{V}$($V$,$d$,$E(B-V)$),BC) [Eq. \ref{bc_flower}]
\item $T_\mathrm{N}$ = f($\log T_\mathrm{eff}$,$\log L/L_{\sun}$) [Eq. \ref{tn}, Table \ref{list_norm}]
\item $[$Z$]$ from isochrones [Table \ref{logteff_grids1}]

\end{enumerate}
The starting value for [Z] is always [Z]$_{\sun}$. 
The transformation from [Z] into [Fe/H] can be done
via the values listed in Table \ref{fe_z}.
If the input parameters $d$, $E(B-V)$, log\,$t$ and [Z] are correctly chosen, within
the errors, the final $<$[Z]$>$ value has to be compatible
with the individual starting value. If not, at least one iteration 
with different starting values has to be performed until no significant changes
are seen. We wish to emphasize that
this intrinsic consistency check makes our method superior to a standard
isochrone fitting technique.

In total, the first iteration for 21 members of the Hyades
gives mean metallicities between 0.029 and 0.032 for logarithmic 
ages between 8.8 and 9.0, respectively. These [Z] values are not
compatible with the starting one. So we applied one iteration with
[Fe/H]\,=\,0.17, which is listed in Dias et al. (2002).

Taking the isochrone for log\,$t$\,=\,8.90 
results in [Z]\,=\,0.028(7), which is the best agreement between 
the starting and the mean value (see Fig. \ref{plot_hyades}). The same 
procedure applying the 
corrected calibration by Alonso et al. (1996) gave [Z]\,=\,0.025(9)
as well as [Z]\,=\,0.028(9), respectively. The corresponding
median values ranged between 0.024 and 0.029 which excellently
agrees with the derived means. 

The age and metallicity is in concordance with the values listed
by Dias et al. (2002).
The test of our method with data for the Hyades gives a satisfying result.
The grid of evolutionary models (Sect. \ref{grids}) and the applied 
calibrations (Sect. \ref{eff_lum}) represent the test data very well.

\begin{figure*}
\begin{center}
\includegraphics[width=162mm]{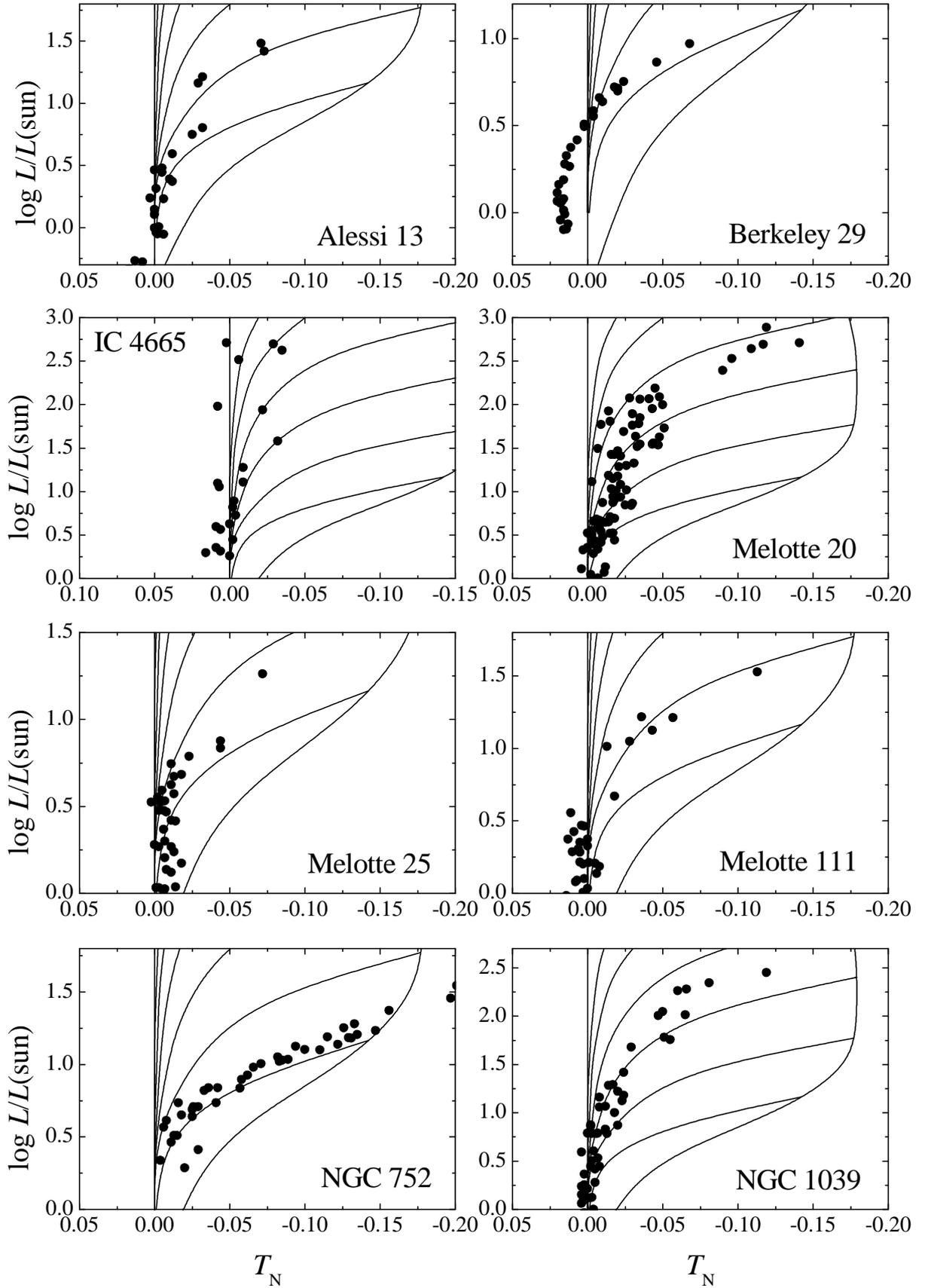}
\caption{The members of investigated clusters in the grid of isochrones. For clarity, only the grid for [Z]\,=\,0.02 is included.}
\label{plot_iso1}
\end{center}
\end{figure*}

\begin{figure*}
\begin{center}
\includegraphics[width=162mm]{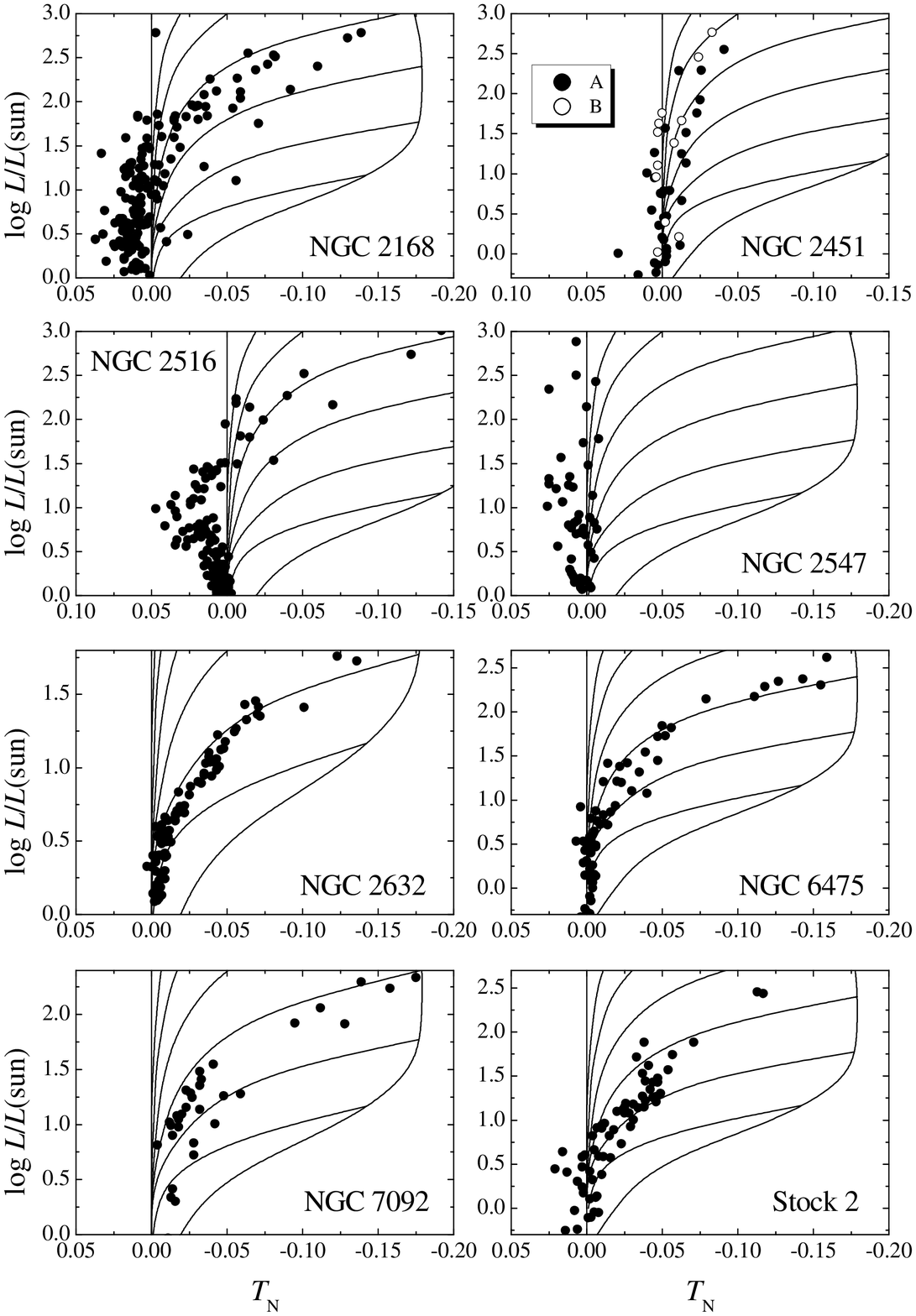}
\caption{Same as Fig. \ref{plot_iso1}.}
\label{plot_iso2}
\end{center}
\end{figure*}

\begin{figure*}
\begin{center}
\includegraphics[width=162mm]{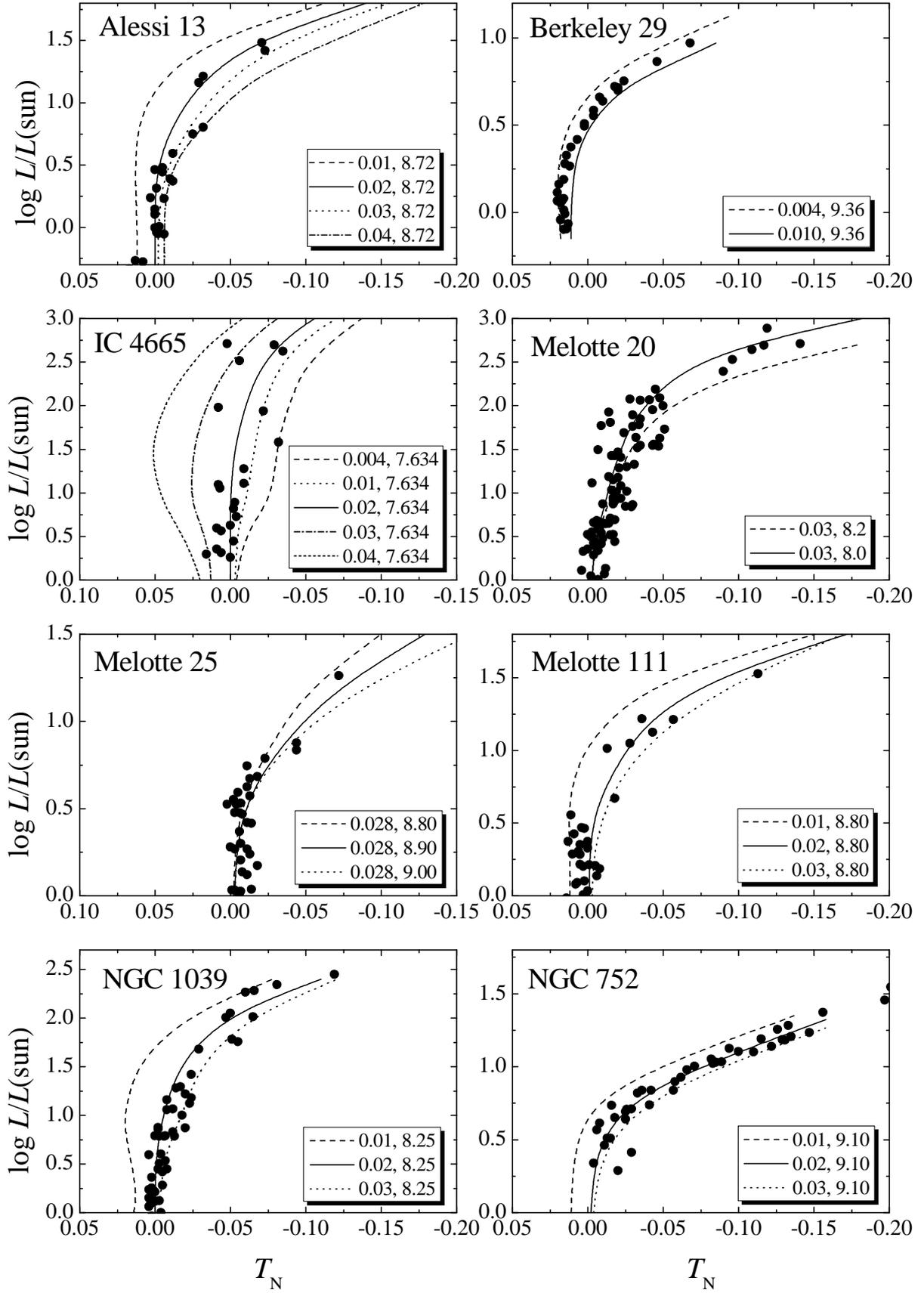}
\caption{The isochrone fits for investigated clusters.}
\label{plot_z1}
\end{center}
\end{figure*}

\begin{figure*}
\begin{center}
\includegraphics[width=162mm]{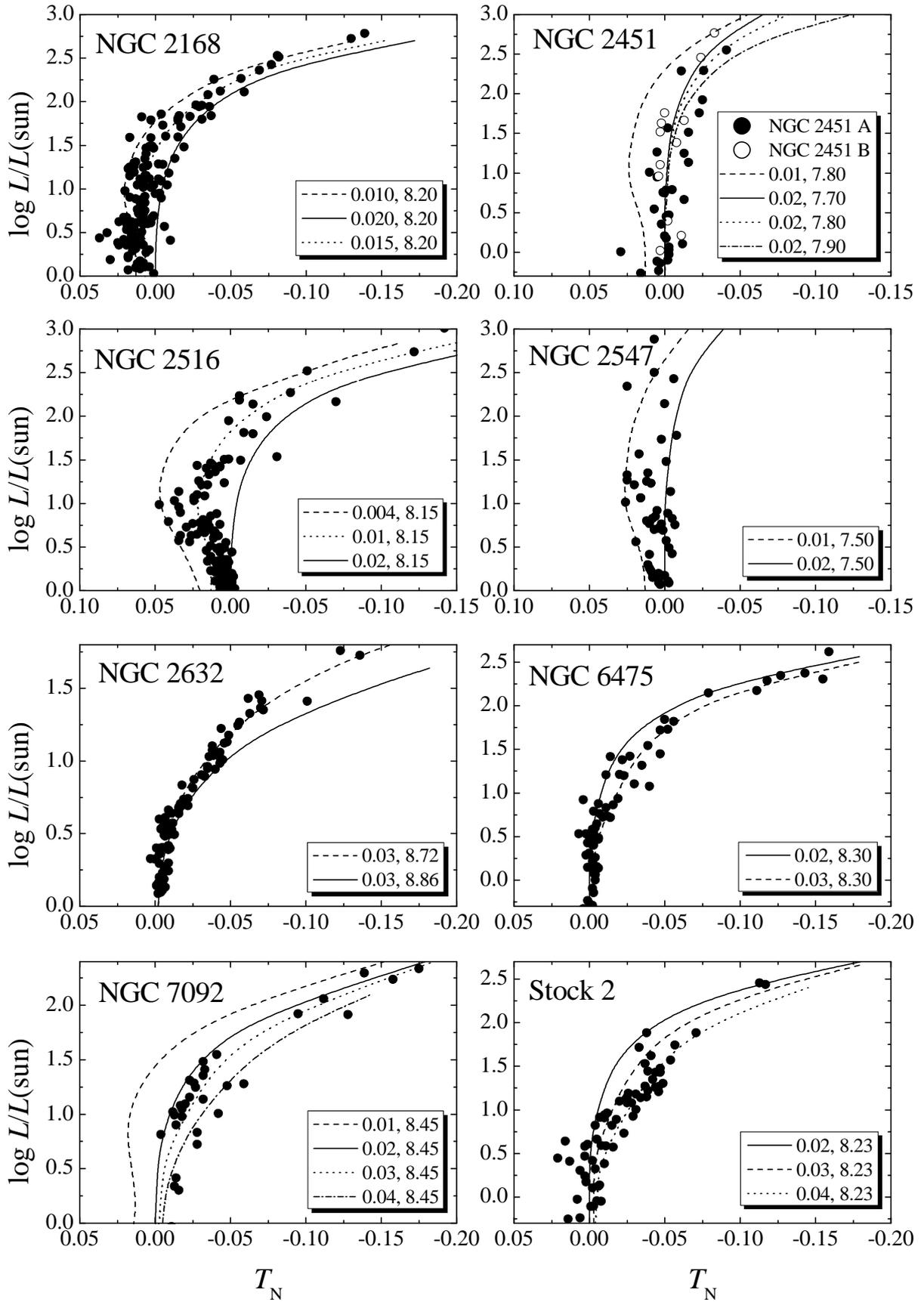}
\caption{Same as Fig. \ref{plot_z1}.}
\label{plot_z2}
\end{center}
\end{figure*}

\subsection{Limitations} \label{limits}

Before we discuss the apparent limitations of our method, we wish to stress again that
the build-in intrinsic consistency check makes our method superior to a standard
isochrone fitting technique. 

On the other hand, there are no tests for the uniqueness
of the final solution incorporated. One needs reliable input parameters
for the age, reddening and distance. Looking at Table \ref{targets_webda}
we find ``iterative changes'' for the distance and reddening of up to 20\%.
Comparing these values with the results of Paunzen \& Netopil (2006), we
conclude that at least 60\% of their evaluated data material fulfil
these requirements. For the age, even a much larger spread can be treated
by our method (see Berkeley 29 and NGC 2168). 

The quality of published photometric data is highly divergent in the
literature (Mermilliod \& Paunzen 2003). In addition to the errors due
to photon noise and the reduction process (for example point-spread-function fitting),
the definition and transformation of the instrumental to a standard 
photometric systems, lacks all consistency much of the time. However, the
high quality checks within WEBDA guarantees that the most divergent
cases can be easily detected and rejected.

The MS of open clusters is shifted by a maximum of $M_\mathrm{V}$\,=\,0.754\,mag
caused by binaries of equal mass. Sharma et al. (2008) deduced a spread for
it of about 0.2\,mag in $(V-I)$. Hurley
\& Tout (1998) simulated the cluster MS with a realistic model of the
binary fraction, concluding that there is a very pronounced ``single star
MS'', a much less pronounced one for a mass fraction equal to one and that in between 
there is only a sparse populated one. The misleading due to this effect can therefore
in a statistically point-of-view be neglected.

One severe problem is the identification and choice of true cluster members. The most
convenient way is to use kinematic data, like the proper motion and radial
velocity for the individual stars. Recent data (Frinchaboy \& 
Majewski 2008, Mermilliod et al. 2009) improved the membership criteria significantly in this respect.
However, these data are mainly available for nearby open clusters. For more distant
clusters, at least two statistical methods were developed to overcome the problem.
Koposov et al. (2008) presented a technique to derive the clusters MS
by counting the number of stars within segments of circles around the apparent
cluster centre and plotting a density colour-magnitude diagram. 
With this method, they were able to detect new open clusters
within the 2MASS survey. The other approach is to perform a statistically cleaning of 
the overall (cluster and field) colour-magnitude diagram with several assumptions
about the stellar content for the line of sight (Piatti et al. 2009). Both methods proved 
to be very robust in a statistical sense. These algorithm could help to 
perform a pre-selection of bona-fide members for distant open clusters before
applying our method to derive the metallicity and improved cluster parameters.
Because our method has to be seen from a statistical point-of-view, the 
choice of ``true'' members may decrease the error, but does not 
influence the average value itself. Looking at Figs. \ref{plot_iso1} and 
\ref{plot_iso2}, a clear and distinct MS can be seen for all open clusters. 
Even if there are apparent non-members at the MS, the result will be not 
statistically affected similar to a Hess-diagramm for isochrone fitting 
(Koposov et al. 2008).

\section{Analysis of additional sixteen open cluster data} \label{analysis}

As a further step, we applied our method to sixteen additional open
clusters (Table \ref{targets_webda}). The photometric data were taken from
WEBDA. Figures \ref{plot_iso1} and 
\ref{plot_iso2} show the cluster data within the evolutionary grids
for [Z]\,=\,0.02 and the isochrone ages according to Fig. \ref{plot_grids}.
The cluster parameters taken from WEBDA, mainly based on the catalogue by Dias et 
al. (2002), are listed in Table \ref{targets_webda}.

The separate isochrone fits are shown in Figs. \ref{plot_z1} and \ref{plot_z2},
which we will now discuss in more detail. The summary of all results are
listed in Table \ref{targets_webda}.

{\it Alessi 13:} The only investigation of this cluster was done by Kharchenko et al. (2005),
who did not include any metallicity value on the basis of $BV$ from the ASCC2.5 catalogue. 
The Hertzsprung-Russell diagram shows a clear MS. Taking their cluster parameters results 
in [Z]\,=\,0.027(7), similar to the Hyades.

{\it Berkeley 29:} With a distance of about 15\,kpc, it is one of the most
distant galactic open clusters known. The papers of Kaluzny (1994) and Tosi et al. (2004)
suggest an higher age than listed in WEBDA. We got a value of [Z]\,=\,0.007(2), which
is in the range of that published by Magrini et al. (2009) deduced from red giant
members. The derived age is between the values published by Dias et al. (2002; 1.06\,Gyr)
and Tosi et al. (2004; 3.4 to 3.7\,Gyr) and fits the isochrones very well.

{\it IC 4665:} The photometric measurements especially for stars with $V$\,$>$\,12.5\,mag
exhibit a significant scatter probably caused by the young age and differential 
reddening. This makes the determination of the MS rather difficult.
We calculated a value of [Z]\,=\,0.022(9).

{\it Melotte 20:} The published metallicity values range from 
$-$0.05\,$<$\,[Fe/H]\,$<$\,+0.07 for this young and near cluster. 
For the first iteration, based on recent $UBV$ CCD photometry,
we chose [Fe/H]\,=\,+0.20, which is supported by a $\delta$($U-B$)$_{0.6}$
of $-$0.03 for the members and the corresponding metallicity calibration from 
Cameron (1985b). The best isochrone fit was found for log\,$t$\,=\,8.05 and 
[Z]\,=\,0.03. The mean values for about 60 stars for log\,$t$\,=\,7.8, 8.0 and 8.2 are 
[Z]\,=\,0.031(7), 0.030(7) as well as 0.027(8), respectively. As described in Sect.
\ref{test_hyades}, the higher metallicity makes an estimation for solar luminosity
stars more uncertain. As for the Hyades, we excluded members with calculated 
[Z]\,$>$\,0.04 values resulting in [Z]\,=\,0.028(7) or [Fe/H]\,=\,+0.18. Melotte 20
seems more metal rich and older than previously thought.

{\it Melotte 111:} This open cluster is well investigated with 
[Fe/H]\,=\,$-$0.05 according to Gratton (2000). The published distances
range from 81\,pc (Makarov 2003) to 96\,pc (Dias et al. 2002). Because
the metallicity is very well known, it is possible to estimate the best
distance value. The colour-magnitude diagram shows a large scatter. Nevertheless, 
it is possible to establish an MS from which we derived $d$\,=\,96\,pc and
[Z]\,=\,0.028(8) as well as $d$\,=\,81\,pc and [Z]\,=\,0.018(8), both with log\,$t$\,=\,8.80, 
respectively. The latter value of [Z] agrees very well with published ones from the 
literature. We therefore conclude that the lower distance value is the more
probable one.

{\it NGC 752:} We used the parameters given by Anthony-Twarog et al. (2006) and 
Taylor (2007) for our analysis. Several publications indicate that NGC 752
exhibits a nearly solar metallicity. We find [Z]\,=\,0.021(5) from data of 40 members.

{\it NGC 1039:} It is one of the clusters for which rather different
metallicity values are published. There is a sufficient
number of observed members to apply our method. Cameron (1985a) listed 
a metallicity of [Fe/H]\,=\,$-$0.291, whereas Schuler et al. (2003) deduced
[Fe/H]\,=\,+0.07(4) from five ``warm stars''. We investigated the isochrones
for [Z]\,=\,0.01, 0.02 and 0.03, which show that the metallicity value
from Cameron (1985a) does not fit the Hertzsprung-Russell diagram. Our final
result of [Z]\,=\,0.023(7) agrees very well with that by Schuler et al. (2003).

{\it NGC 2168:} With a distance of approximately 820\,pc from the Sun, NGC 2168
is the most distant cluster apart from Berkeley 29 of our target sample. It is a 
perfect test case
for more distant open clusters because we find diverging published values ranging
from 816\,pc (Kharchenko et al. 2005) 
to 918\,pc (Kalirai et al. 2003). Barrado y Navascues et al. (2001) list
[Fe/H]\,=\,$-$0.21, whereas we calculated [Z]\,=\,0.014(3) for the best isochrone
fit for log\,$t$\,=\,8.20.

{\it NGC 2451 A and B:} Maitzen \& Catalano (1986) were the first who suggested 
that NGC 2451 actually comprises of two separate clusters, which happen to lie
in a common line of sight. This conclusion was proved by R{\"o}ser \& Bastian (1994)
on the basis of kinematic data. H{\"u}nsch et al. (2003) estimated an age of 50 and
80\,Myr, a distance of 206 and 370\,pc and a reddening of 0.01 and 0.05\,mag for the 
individual components. The rather low metallicity value of [Fe/H]\,=\,$-$0.50 published
by Lyng\aa\,\,\& Wramdemark (1984) does not take into account the ``binary nature'' of this
cluster and is most certainly incorrect (H{\"u}nsch et al. 2003).
We included the data of both components in one plot to show the
similar metallicities and the rather young ages. Our metallicity estimation is
[Z]\,=\,0.020(5) and [Z]\,=\,0.019(4) for NGC 2451 A and B, respectively. 

{\it NGC 2516:} The large number of members and peculiar stars made NGC 2516
a target of several detailed investigations. We used the cluster parameters
listed by Sung et al. (2002) who included all up-to-date published data
and new optical as well as X-ray observations. We found a best
fit for log\,$t$\,=\,8.15 and derived [Z]\,=\,0.015(5), which is compatible 
with the value mentioned before. 

{\it NGC 2547:} We compared the cluster parameters published by 
Dias et al. (2002), Naylor et al. (2002), Lyra et al. (2006) and 
Kharchenko et al. (2005) using the appropriate isochrones. The best fit
was obtained with the values from Naylor et al. (2002), which were used
to calculate [Z]\,=\,0.017(5). 

{\it NGC 2632:} For Praesepe we found several metallicity estimates in the
literature ranging from +0.04 to +0.19 (Gratton 2000). For the reddening, we
used $E(B-V)$\,=\,0.027\,mag (Taylor 2006), which fits the photometric data
much better than the value (0.009) listed by Dias et al. (2001). The latter
resulted in effective temperatures which were systematically two to three 
percent too low. For an age of log\,$t$\,=\,8.72 (An et al. 2007), we
get [Z]\,=\,0.028(6) which means that the metal content of Praesepe is similar 
to that of the Hyades.

{\it NGC 6475:} Meynet et al. (1993) list this cluster in their standard compilation
with an age of log\,$t$\,=\,8.35, which is younger than the age (8.475) given 
in Dias et al. (2002). We found a best fit for log\,$t$\,=\,8.30 yielding
[Z]\,=\,0.028(7).

{\it NGC 7092:} The published cluster parameters by Robichon et al. (1999),
Dias et al. (2002) and Kharchenko et al. (2005) agree within the errors very
well, resulting in [Z]\,=\,0.026(6).

{\it Stock 2:} The cluster parameters by Dias et al. (2002, Table \ref{targets_webda})
do not coincide with those by Kharchenko et al. (2005), who list a larger distance 
of 380\,pc and a smaller reddening of 0.34\,mag, but a comparable age.
We checked both sets of cluster parameters and found a much better agreement using
the ones by Kharchenko et al. (2005), which we took to estimate [Z]\,=\,0.032(10).

In Table \ref{targets_webda} we summarize our results and compare them with the
cluster parameters from WEBDA. In addition, we list the
[Fe/H] values either from Magrini et al. (2009) based on high resolution spectroscopy 
or Chen et al. (2003), which is a mean of the at that time published values. 
The latter did not examine the data individually to 
see whether there were important differences among the clusters in the different
catalogues. So the values can only serve as a guideline.

From the complete sample of seventeen open clusters, we deduced a mean error
of $\sigma$[Z]\,=\,0.006 for the derived metallicities. Averaging the errors
listed in Magrini et al. (2009) results in $\sigma$[Z]\,=\,0.004.
Therefore our method provides in a statistical sense a comparable error level,
to that of high resolution spectroscopy.

\section{Conclusion and outlook}

The situation of a homogeneous metallicity determination of open
clusters is still very unsatisfying. Recent compilations in this respect
suffer from the bias introduced by averaging values from many different sources 
and applied techniques (e.g. isochrone
fitting and spectroscopic determinations) as well as large uncertainties 
for single object estimations which yield a significant error of the means. 

The overall metallicity of open clusters and their members is an important
astrophysical parameter for the understanding and modelling of the formation and
evolution from a local (stars, binary systems, etc.) and global (the Milky Way
as Galaxy) point of view. The key observational constraint of the Galactic 
metallicity gradient, for example, is still very inaccurate. Results of
open clusters should be compared to abundances determinations of star
groups like Cepheids, or globular clusters to get a consistent 
and global picture.

We presented a statistically robust method to determine the cluster metallicity
in an iterative way. With a rough estimate of the age, reddening and distance
of the cluster, the algorithm iterates within normalized evolutionary grids 
(Schaller et al. 1992, Charbonnel et al. 1993, Schaerer et al. 1993, 
and Claret \& Gimenez 1998) to find the best numerical fit of all free four cluster 
parameters. As input data, Johnson $BV$ measurements of members defining the 
MS of the individual cluster are needed. These data are then 
calibrated using a corrected
effective calibration by Alonso et al. (1996) and the bolometric correction
by Flower (1996). One major advantage of the proposed method is the statistical treatment
of many objects simultaneously, which minimizes individual outliers as well as
erroneous measurements. 

Our method was demonstrated and tested with data from the Hyades yielding 
an excellent agreement with the published values. Using the photometric data
from WEBDA, we have selected sixteen additional open clusters, namely Alessi 13, 
Berkeley 29, IC 4665, 
Melotte 20, Melotte 111 , NGC 752 , NGC 1039, NGC 2168, NGC 2451\,A, NGC 2451\,B,
NGC 2516, NGC 2547, NGC 2632, NGC 6475, NGC 7092 and Stock 2,
within a distance of 1000\,pc (exception: Berkeley 29 with 15\,kpc)
around the Sun and applied our method to determine the overall
mean metallicity. In addition, the published cluster parameters were checked and, if
necessary, improved. 

The algorithm can be employed in a semi-automatic way for a much
larger sample of open clusters and can be extended to any other photometric system.
As a next step, we will apply the algorithm to the data included in WEBDA to
determine cluster metallicities for a statistical significant number of aggregates.

\begin{acknowledgements}
We thank our colleagues Christian St{\"u}tz and Martin Netopil for the help 
as well as the referee for very valuable comments.
This work was supported by the financial contributions of the Austrian Agency for International 
Cooperation in Education and Research (WTZ CZ-11/2008) and of the City of Vienna 
(Hochschuljubil{\"a}umsstiftung project: H-1930/2008). 
\end{acknowledgements}

\begin{appendix}
\section{Evolutionary grids}
Here we list the complete evolutionary grids from Claret \& Gimenez (1998), 
Schaller et al. (1992), Charbonnel et al. (1993) and Schaerer et al. (1993)
used for this paper. They are normalized using the ZAMS model for [0.70:0.28:0.02],
listed in Table \ref{list_norm}, to the logarithmic effective temperature 
$T_\mathrm{N}$. 
\begin{table*}
\begin{center}
\caption[]{The complete evolutionary grids in time steps of $\Delta$log\,$t$\,=\,0.2 used for this study.
Within the individual isochrones, a linear interpolation was performed.}
\begin{tabular}{cccccc|cccccc}
\hline
\hline
\multicolumn{6}{l|}{ZAMS} & \multicolumn{6}{l}{log\,$t$\,=\,7.2} \\	  
Z &	0.004 & 0.01 & 0.02 & 0.03 & 0.04 & Z &	0.004 & 0.01 & 0.02 & 0.03 & 0.04 \\
$\log L/L_{\sun}$ & \multicolumn{5}{c|}{$T_\mathrm{N}$} & $\log L/L_{\sun}$ & \multicolumn{5}{c}{$T_\mathrm{N}$} \\
$-$0.3 &  & +0.020 & +0.000 & $-$0.001 &  \\
+0.0 & +0.019 & +0.013 & +0.000 & $-$0.003 & $-$0.004 & +0.0 & +0.019 & +0.013 & +0.000 & $-$0.003 & $-$0.004 \\ 
+0.3 & +0.026 & +0.014 & +0.000 & $-$0.004 & $-$0.007 & +0.3 & +0.026 & +0.014 & +0.000 & $-$0.004 & $-$0.007 \\
+0.6 & +0.035 & +0.020 & +0.000 & $-$0.005 & $-$0.013 & +0.6 & +0.035 & +0.019 & +0.000 & $-$0.005 & $-$0.013 \\ 
+0.9 & +0.042 & +0.027 & +0.000 & $-$0.008 & $-$0.024 & +0.9 & +0.045 & +0.026 & +0.000 & $-$0.009 & $-$0.025 \\
+1.2 & +0.049 & +0.028 & +0.000 & $-$0.010 & $-$0.027 & +1.2 & +0.049 & +0.027 & +0.000 & $-$0.012 & $-$0.028 \\ 
+1.5 & +0.054 & +0.028 & +0.000 & $-$0.011 & $-$0.027 & +1.5 & +0.053 & +0.027 & $-$0.001 & $-$0.013 & $-$0.029 \\
+1.8 & +0.056 & +0.027 & +0.000 & $-$0.011 & $-$0.027 & +1.8 & +0.053 & +0.026 & $-$0.002 & $-$0.014 & $-$0.030 \\ 
+2.1 & +0.055 & +0.026 & +0.000 & $-$0.012 & $-$0.027 & +2.1 & +0.049 & +0.022 & $-$0.004 & $-$0.015 & $-$0.031 \\ 
+2.4 & +0.050 & +0.025 & +0.000 & $-$0.011 & $-$0.026 & +2.4 & +0.043 & +0.017 & $-$0.007 & $-$0.017 & $-$0.034 \\ 
+2.7 & +0.046 & +0.023 & +0.000 & $-$0.011 & $-$0.024 & +2.7 & +0.035 & +0.011 & $-$0.011 & $-$0.022 & $-$0.043 \\ 
+3.0 & +0.044 & +0.022 & +0.000 & $-$0.010 & $-$0.021 & +3.0 & +0.026 & +0.003 & $-$0.018 & $-$0.029 & $-$0.048 \\ 
+3.3 & & +0.021 & +0.000 & $-$0.010 &  & +3.3 & & $-$0.007 & $-$0.028 & $-$0.038 &  \\ 
+3.6 & & +0.019 & +0.000 & $-$0.009 &  & +3.6 & & $-$0.021 & $-$0.042 & $-$0.053 &  \\
\hline
\multicolumn{6}{l|}{log\,$t$\,=\,7.4} & \multicolumn{6}{l}{log\,$t$\,=\,7.6} \\ 
Z &	0.004 & 0.01 & 0.02 & 0.03 & 0.04 & Z &	0.004 & 0.01 & 0.02 & 0.03 & 0.04 \\
$\log L/L_{\sun}$ & \multicolumn{5}{c|}{$T_\mathrm{N}$} & $\log L/L_{\sun}$ & \multicolumn{5}{c}{$T_\mathrm{N}$} \\
& & & & & & $-$0.3 & & +0.020 & +0.000 &  \\
+0.0 & +0.019 & +0.013 & +0.000 & $-$0.003 & $-$0.004 & +0.0 & +0.019 & +0.013 & +0.000 & $-$0.003 & $-$0.004 \\ 
+0.3 & +0.026 & +0.014 & +0.000 & $-$0.004 & $-$0.007 & +0.3 & +0.026 & +0.014 & +0.000 & $-$0.004 & $-$0.007 \\ 
+0.6 & +0.035 & +0.019 & +0.000 & $-$0.005 & $-$0.013 & +0.6 & +0.035 & +0.019 & +0.000 & $-$0.005 & $-$0.013 \\ 
+0.9 & +0.045 & +0.026 & +0.000 & $-$0.010 & $-$0.025 & +0.9 & +0.045 & +0.026 & +0.000 & $-$0.010 & $-$0.025 \\
+1.2 & +0.049 & +0.027 & $-$0.001 & $-$0.012 & $-$0.028 & +1.2 & +0.049 & +0.026 & $-$0.001 & $-$0.013 & $-$0.029 \\ 
+1.5 & +0.053 & +0.026 & $-$0.002 & $-$0.013 & $-$0.029 & +1.5 & +0.052 & +0.024 & $-$0.003 & $-$0.015 & $-$0.031 \\
+1.8 & +0.051 & +0.024 & $-$0.003 & $-$0.015 & $-$0.031 & +1.8 & +0.049 & +0.021 & $-$0.006 & $-$0.018 & $-$0.034 \\
+2.1 & +0.046 & +0.019 & $-$0.007 & $-$0.018 & $-$0.034 & +2.1 & +0.042 & +0.014 & $-$0.011 & $-$0.023 & $-$0.040 \\
+2.4 & +0.038 & +0.013 & $-$0.011 & $-$0.022 & $-$0.041 & +2.4 & +0.031 & +0.006 & $-$0.017 & $-$0.029 & $-$0.046 \\
+2.7 & +0.028 & +0.004 & $-$0.018 & $-$0.029 & $-$0.045 & +2.7 & +0.018 & $-$0.007 & $-$0.029 & $-$0.041 & $-$0.059 \\
+3.0 & +0.016 & $-$0.008 & $-$0.029 & $-$0.040 & $-$0.057 & +3.0 & $-$0.002 & $-$0.024 & $-$0.048 & $-$0.061 & $-$0.079 \\
+3.3 & & $-$0.022 & $-$0.045 & $-$0.056 &  & +3.3 & & $-$0.050 & $-$0.077 & $-$0.091 &  \\
+3.6 & & $-$0.047 & $-$0.072 & $-$0.087 &  & +3.6 & & $-$0.104 & $-$0.139 & $-$0.162 &  \\
\multicolumn{6}{l|}{} & TAMS \\				
& & & & & & $\log L/L_{\sun}$ &  & +3.769 & +3.702 & +3.642 & \\
& & & & & & $T_\mathrm{N}$ &  & $-$0.153 & $-$0.168 & $-$0.171 \\
& & & & & & $\log T_\mathrm{eff}$ &  & +4.252 & +4.226 & +4.213 & \\
\hline
\multicolumn{6}{l|}{log\,$t$\,=\,7.8} & \multicolumn{6}{l}{log\,$t$\,=\,8.0} \\ 
Z &	0.004 & 0.01 & 0.02 & 0.03 & 0.04 & Z &	0.004 & 0.01 & 0.02 & 0.03 & 0.04 \\
$\log L/L_{\sun}$ & \multicolumn{5}{c|}{$T_\mathrm{N}$} & $\log L/L_{\sun}$ & \multicolumn{5}{c}{$T_\mathrm{N}$} \\ 
$-$0.3 &  & +0.020 & +0.000 & +0.001 &  & $-$0.3 &  & +0.020 & +0.000 & +0.001 &  \\
+0.0 & +0.023 & +0.013 & +0.000 & $-$0.003 & $-$0.004 & +0.0 & +0.023 & +0.013 & +0.000 & $-$0.003 & $-$0.005 \\
+0.3 & +0.028 & +0.014 & +0.000 & $-$0.004 & $-$0.007 & +0.3 & +0.028 & +0.014 & +0.000 & $-$0.004 & $-$0.007 \\
+0.6 & +0.034 & +0.019 & +0.000 & $-$0.005 & $-$0.013 & +0.6 & +0.035 & +0.018 & $-$0.001 & $-$0.005 & $-$0.013 \\
+0.9 & +0.047 & +0.025 & $-$0.001 & $-$0.011 & $-$0.026 & +0.9 & +0.047 & +0.024 & $-$0.002 & $-$0.012 & $-$0.026 \\
+1.2 & +0.053 & +0.024 & $-$0.002 & $-$0.014 & $-$0.030 & +1.2 & +0.052 & +0.022 & $-$0.004 & $-$0.016 & $-$0.032 \\
+1.5 & +0.052 & +0.022 & $-$0.005 & $-$0.018 & $-$0.033 & +1.5 & +0.048 & +0.018 & $-$0.009 & $-$0.022 & $-$0.038 \\
+1.8 & +0.045 & +0.017 & $-$0.009 & $-$0.022 & $-$0.039 & +1.8 & +0.039 & +0.011 & $-$0.015 & $-$0.029 & $-$0.047 \\
+2.1 & +0.035 & +0.008 & $-$0.017 & $-$0.030 & $-$0.048 & +2.1 & +0.025 & $-$0.003 & $-$0.029 & $-$0.043 & $-$0.066 \\
+2.4 & +0.021 & $-$0.005 & $-$0.029 & $-$0.041 & $-$0.060 & +2.4 & +0.004 & $-$0.023 & $-$0.048 & $-$0.063 & $-$0.095 \\ 
+2.7 & +0.000 & $-$0.024 & $-$0.048 & $-$0.062 & $-$0.091 & +2.7 & $-$0.026 & $-$0.055 & $-$0.085 & $-$0.104 & $-$0.129 \\ 
+3.0 & $-$0.036 & $-$0.054 & $-$0.082 & $-$0.100 & $-$0.128 & +3.0 & $-$0.092 & $-$0.126 & $-$0.165 &  &  \\ 
+3.3 &  & $-$0.113 & $-$0.144 & $-$0.173 &  &  &  \\
TAMS & \multicolumn{5}{l|}{} & TAMS \\	 
$\log L/L_{\sun}$ & +3.447 & +3.421 & +3.371 & +3.317 & +3.222 & $\log L/L_{\sun}$ & +3.086 & +3.076 & +3.039 & +2.987 & +2.937 \\ 
$T_\mathrm{N}$ & $-$0.117 & $-$0.151 & $-$0.170 & $-$0.179 & $-$0.180 & $T_\mathrm{N}$ & $-$0.113 & $-$0.152 & $-$0.173 & $-$0.182 & $-$0.176 \\ 
$\log T_\mathrm{eff}$ & +4.239 & +4.196 & +4.168 & +4.132 & +4.141 & $\log T_\mathrm{eff}$ & +4.181 & +4.140 & +4.110 & +4.092 & +4.090 \\
\hline
\end{tabular}
\label{logteff_grids1}
\end{center}
\end{table*}

\begin{table*}
\begin{center}
\addtocounter{table}{-1}
\caption[]{continued}
\begin{tabular}{cccccc|cccccc}
\hline
\hline
\multicolumn{6}{l|}{log\,$t$\,=\,8.2} & \multicolumn{6}{l}{log\,$t$\,=\,8.4} \\	  
Z &	0.004 & 0.01 & 0.02 & 0.03 & 0.04 & Z &	0.004 & 0.01 & 0.02 & 0.03 & 0.04 \\
$\log L/L_{\sun}$ & \multicolumn{5}{c|}{$T_\mathrm{N}$} & $\log L/L_{\sun}$ & \multicolumn{5}{c}{$T_\mathrm{N}$} \\
$-$0.3 &  & +0.020 & +0.000 & +0.001 &  & $-$0.3 & & +0.020 & +0.000 & +0.001 &  \\ 
+0.0 & +0.019 & +0.013 & +0.000 & $-$0.003 & $-$0.005 & +0.0 & +0.019 & +0.013 & +0.000 & $-$0.003 & $-$0.005 \\
+0.3 & +0.027 & +0.013 & +0.000 & $-$0.004 & $-$0.006 & +0.3 & +0.027 & +0.013 & $-$0.001 & $-$0.004 & $-$0.007 \\ 
+0.6 & +0.036 & +0.017 & $-$0.001 & $-$0.007 & $-$0.013 & +0.6 & +0.037 & +0.017 & $-$0.002 & $-$0.006 & $-$0.013 \\ 
+0.9 & +0.046 & +0.022 & $-$0.003 & $-$0.014 & $-$0.026 & +0.9 & +0.047 & +0.019 & $-$0.006 & $-$0.017 & $-$0.029 \\ 
+1.2 & +0.047 & +0.019 & $-$0.007 & $-$0.020 & $-$0.034 & +1.2 & +0.047 & +0.015 & $-$0.013 & $-$0.026 & $-$0.042 \\ 
+1.5 & +0.041 & +0.013 & $-$0.014 & $-$0.028 & $-$0.045 & +1.5 & +0.040 & +0.003 & $-$0.025 & $-$0.040 & $-$0.054 \\ 
+1.8 & +0.031 & +0.000 & $-$0.027 & $-$0.041 & $-$0.059 & +1.8 & +0.028 & $-$0.019 & $-$0.047 & $-$0.063 & $-$0.081 \\
+2.1 & +0.010 & $-$0.023 & $-$0.049 & $-$0.064 & $-$0.090 & +2.1 & +0.009 & $-$0.058 & $-$0.090 & $-$0.108 & $-$0.130 \\ 
+2.4 & $-$0.043 & $-$0.057 & $-$0.087 & $-$0.106 & $-$0.143 & +2.4 & $-$0.088 & $-$0.142 & $-$0.179 & \\ 
+2.7 & $-$0.101 & $-$0.132 & $-$0.172 &  &  &  \\ 
TAMS & \multicolumn{5}{l|}{} & TAMS \\
$\log L/L_{\sun}$ & +2.762 & +2.771 & +2.729 & +2.663 & +2.613 &  $\log L/L_{\sun}$ & +2.464 & +2.442 & +2.402 & +2.343 & +2.334 \\ 
$T_\mathrm{N}$ & $-$0.113 & $-$0.154 & $-$0.178 & $-$0.185 & $-$0.180 & $T_\mathrm{N}$ & $-$0.113 & $-$0.154 & $-$0.179 & $-$0.186 & $-$0.182 \\
$\log T_\mathrm{eff}$ & +4.126 & +4.083 & +4.052 & +4.033 & +4.032 &  $\log T_\mathrm{eff}$ & +4.074 & +4.025 &+ 3.993 & +3.975 & +3.980 \\
\hline
\multicolumn{6}{l|}{log\,$t$\,=\,8.6} & \multicolumn{6}{l}{log\,$t$\,=\,8.8} \\	  
Z &	0.004 & 0.01 & 0.02 & 0.03 & 0.04 & Z &	0.004 & 0.01 & 0.02 & 0.03 & 0.04 \\
$\log L/L_{\sun}$ & \multicolumn{5}{c|}{$T_\mathrm{N}$} & $\log L/L_{\sun}$ & \multicolumn{5}{c}{$T_\mathrm{N}$} \\ 
$-$0.3 &  & +0.020 & $-$0.001 & +0.001 &  & $-$0.3 &  & +0.021 & $-$0.002 & +0.001 & \\
+0.0 & +0.019 & +0.012 & $-$0.001 & $-$0.003 & $-$0.006 & +0.0 & +0.020 & +0.012 & $-$0.001 & $-$0.003 & $-$0.006 \\ 
+0.3 & +0.027 & +0.013 & $-$0.001 & $-$0.004 & $-$0.007 & +0.3 & +0.028 & +0.012 & $-$0.001 & $-$0.004 & $-$0.007 \\ 
+0.6 & +0.035 & +0.015 & $-$0.003 & $-$0.008 & $-$0.015 & +0.6 & +0.035 & +0.013 & $-$0.005 & $-$0.011 & $-$0.016 \\ 
+0.9 & +0.043 & +0.019 & $-$0.010 & $-$0.023 & $-$0.035 & +0.9 & +0.038 & +0.008 & $-$0.019 & $-$0.032 & $-$0.042 \\ 
+1.2 & +0.033 & +0.006 & $-$0.022 & $-$0.037 & $-$0.055 & +1.2 & +0.020 & $-$0.011 & $-$0.039 & $-$0.056 & $-$0.077 \\ 
+1.5 & +0.012 & $-$0.015 & $-$0.043 & $-$0.061 & $-$0.090 & +1.5 & $-$0.017 & $-$0.050 & $-$0.083 & $-$0.104 & $-$0.137 \\
+1.8 & $-$0.032 & $-$0.056 & $-$0.088 & $-$0.107 & $-$0.137 & +1.8 & $-$0.040 & $-$0.151 & \\ 
+2.1 & $-$0.100 & $-$0.150 & \multicolumn{3}{l|}{} \\ 
TAMS & \multicolumn{5}{l|}{} & TAMS \\ 
$\log L/L_{\sun}$ & +2.135 & +2.118 & +2.082 & +2.040 & +1.981 &  $\log L/L_{\sun}$ & +1.845 & +1.802 & +1.772 & +1.736 & +1.671 \\
$T_\mathrm{N}$ & $-$0.111 & $-$0.154 & $-$0.179 & $-$0.186 & $-$0.183 & $T_\mathrm{N}$ & $-$0.106 & $-$0.153 & $-$0.178 & $-$0.185 & $-$0.177 \\
$\log T_\mathrm{eff}$ & +4.017 & +3.967 & +3.936 & +3.921 & +3.918 & $\log T_\mathrm{eff}$ & +3.956 & +3.922 & +3.881 & +3.866 & +3.867 \\
\hline
\multicolumn{6}{l|}{log\,$t$\,=\,9.0} & \multicolumn{6}{l}{log\,$t$\,=\,9.2} \\	  
Z &	0.004 & 0.01 & 0.02 & 0.03 & 0.04 & Z &	0.004 & 0.01 & 0.02 & 0.03 & 0.04 \\
$\log L/L_{\sun}$ & \multicolumn{5}{c|}{$T_\mathrm{N}$} & $\log L/L_{\sun}$ & \multicolumn{5}{c}{$T_\mathrm{N}$} \\ 
$-$0.3 &  & +0.019 & $-$0.002 & +0.001 &  & $-$0.3 &  & +0.019 & $-$0.003 & +0.001 \\
+0.0 & +0.022 & +0.011 & $-$0.001 & $-$0.003 & $-$0.007 & +0.0 & +0.019 & +0.011 & $-$0.002 & $-$0.004 & $-$0.008 \\
+0.3 & +0.025 & +0.011 & $-$0.002 & $-$0.005 & $-$0.008 & +0.3 & +0.022 & +0.009 & $-$0.005 & $-$0.008 & $-$0.011 \\
+0.6 & +0.028 & +0.008 & $-$0.009 & $-$0.016 & $-$0.023 & +0.6 & +0.020 & +0.001 & $-$0.017 & $-$0.025 & $-$0.034 \\
+0.9 & +0.021 & $-$0.005 & $-$0.033 & $-$0.047 & $-$0.061 &  +0.9 & $-$0.012 & $-$0.032 & $-$0.061 & $-$0.075 & $-$0.095 \\
+1.2 & $-$0.011 & $-$0.044 & $-$0.076 & $-$0.096 & $-$0.133 & +1.2 & $-$0.079 & \\ 
+1.5 & $-$0.090 & $-$0.145 & \multicolumn{3}{l|}{} \\ 
TAMS & \multicolumn{5}{l|}{} & TAMS \\
$\log L/L_{\sun}$ & +1.533 & +1.516 & +1.478 & +1.411 & +1.351 & $\log L/L_{\sun}$ & +1.242 & +1.197 & +1.165 & +1.120 & +1.057 \\
$T_\mathrm{N}$ & $-$0.099 & $-$0.149 & $-$0.171 & $-$0.173 & $-$0.167 & $T_\mathrm{N}$ & $-$0.090 & $-$0.128 & $-$0.145 & $-$0.143 & $-$0.134 \\
$\log L/L_{\sun}$ & +3.919 & +3.863 & +3.834 & +3.819 & +3.817 & $\log L/L_{\sun}$ & +3.873 & +3.824 & +3.800 & +3.793 & +3.793 \\ 
\hline
\multicolumn{6}{l|}{log\,$t$\,=\,9.4} & \multicolumn{6}{l}{log\,$t$\,=\,9.6} \\	  
Z &	0.004 & 0.01 & 0.02 & 0.03 & 0.04 & Z &	0.004 & 0.01 & 0.02 & 0.03 & 0.04 \\
$\log L/L_{\sun}$ & \multicolumn{5}{c|}{$T_\mathrm{N}$} & $\log L/L_{\sun}$ & \multicolumn{5}{c}{$T_\mathrm{N}$} \\ 
+0.0 & +0.019 & +0.011 & $-$0.003 & $-$0.006 & $-$0.011 & +0.0 & +0.018 & +0.010 & $-$0.004 & $-$0.008 & $-$0.014 \\
+0.15 & +0.019 & +0.011 & $-$0.005 & $-$0.008 & $-$0.013 & +0.15 & +0.017 & +0.008 & $-$0.008 & $-$0.014 & $-$0.020 \\
+0.3 & +0.018 & +0.007 & $-$0.008 & $-$0.012 & $-$0.017 & +0.3 & +0.015 & +0.002 & $-$0.016 & $-$0.028 & $-$0.037 \\
+0.45 & +0.014 & +0.001 & $-$0.014 & $-$0.022 & $-$0.029 & +0.45 & +0.008 & $-$0.010 & $-$0.029 & $-$0.046 \\
+0.6 & +0.008 & $-$0.012 & $-$0.030 & $-$0.040 & $-$0.056 &  +0.6 & $-$0.018 & $-$0.040 & $-$0.060  \\
+0.75 & $-$0.004 & $-$0.034 & $-$0.056 & $-$0.070 & $-$0.090 \\
+0.9 & $-$0.028 & $-$0.075 & $-$0.098 & \multicolumn{2}{l|}{} \\
TAMS & \multicolumn{5}{l|}{} & TAMS \\
$\log L/L_{\sun}$ & +0.982 & +0.940 & +0.908 & +0.855 & +0.778 & $\log L/L_{\sun}$ & +0.695 & +0.650 & +0.603 & +0.538 & +0.446 \\
$T_\mathrm{N}$ & $-$0.075 & $-$0.090 & $-$0.103 & $-$0.103 & $-$0.097 & $T_\mathrm{N}$ & $-$0.055 & $-$0.060 & $-$0.064 & $-$0.064 & $-$0.060 \\
$\log L/L_{\sun}$ & +3.839 & +3.808 & +3.785 & +3.781 & +3.778 & $\log L/L_{\sun}$ & +3.818 & +3.795 & +3.774 & +3.771 & +3.763 \\ 
\hline
\end{tabular}
\label{logteff_grids2}
\end{center}
\end{table*}
\end{appendix}
\end{document}